\documentclass[10pt,aps,pre,preprint,showpacs]{revtex4-1}

\usepackage{graphicx}

\begin{document}

\title{Fluctuation Analysis of the Atmospheric Energy Cycle}

\author{Richard Blender}
\affiliation{Meteorological Institute, University of Hamburg, 
Hamburg, Germany}
\email[]{richard.blender@uni-hamburg.de}

\author{Denny Gohlke}
\affiliation{Meteorological Institute, University of Hamburg, 
Hamburg, Germany}
\email[]{Denny.Gohlke@uni-hamburg.de}	

\author{Frank Lunkeit}
\affiliation{Meteorological Institute, University of Hamburg, 
Hamburg, Germany}
\email[]{Frank.Lunkeit@uni-hamburg.de}	

\date{\today}

\begin{abstract}
The atmosphere gains available potential energy by solar radiation and dissipates kinetic energy mainly in the atmospheric boundary layer. We analyze the fluctuations of the global mean energy cycle defined by Lorenz (1955) in a simulation with a simplified hydrostatic model. The energy current densities are well approximated by the generalized Gumbel distribution (Bramwell, Holdsworth and Pinton,  1998) and the Generalized Extreme Value (GEV) distribution. In an attempt to assess the fluctuation relation of Evans, Cohen, and Morriss (1993) we define entropy production by the injected power and use the GEV location parameter as a reference state. The fluctuation ratio reveals a linear behavior in a finite range.
\end{abstract}


\maketitle

\section{Introduction}



The global atmosphere is a physical system driven to a non-equilibrium state by radiative forcing and friction in the atmospheric boundary layer. A well-known diagnostic scheme for the energy flow is the Lorenz energy cycle (LEC) \cite{Lorenz:1955} which includes the zonal mean and the eddy parts of the available potential and the kinetic energies and determines the injected power, the dissipated energy and internal conversions. The LEC constitutes a network of energy currents and can be considered as an atmospheric energy cascade. For the ocean an analogous cycle can be defined \cite{Storch:2012}. The means in the LEC constitute the climate from a dynamical point of view and the fluctuations are related to climate variability.  Note that the properties of the climatological LEC are known from the output of models only, mainly in the reanalysis datasets ERA and NCEP \cite{Li:2007}.  


Our data are produced in a simulation with the atmospheric model PUMA (Portable University Model of the Atmosphere, University of Hamburg), which is a dynamical core based on the hydrostatic primitive equations implemented in complex weather and climate models \cite{Fraedrich:2005}. PUMA is subject to linear surface friction and hyper-diffusion. 
The model is driven by a temperature relaxation towards a steady state close to observations. The neglect of complex parameterizations is outweighed by transparent physical equations and a high numerical efficiency.


Few results for fluctuations in non-equilibrium systems are known. A remarkable finding was that the fluctuation of global observables can be approximated by the generalized Gumbel distribution \cite{Bramwell:1998, Bramwell:2000} which depends on a parameter $k$ denoting the order of the maximum. This parameter was identified as $k \approx \pi/2$, hence a non-integer between the first and the second maximum. A special form of the gamma distribution (the chi-square-distribution) has been fitted to the kinetic energy and the dissipation rate in a spring-block model \cite{Aumaitre:2001}. Since different types of complex systems show the generalized Gumbel distribution, a common origin can be assumed. Hypotheses for the occurrence of this distribution are self-similarity, extremal processes, and  correlations \cite{Dahlstedt:2001,Bertin:2005}. Since the energy currents in the LEC are global averages and the turbulent atmosphere is highly correlated it is worthwhile to test whether the fluctuations follow this distribution.


The Fluctuation Theorem (FT) \cite{Evans:1993,Gallavotti_PRL:1995,Gallavotti_JSP:1995,Evans:2002,Sevick:2008} relates the probabilities of negative and positive entropy productions in non-equilibrium physical systems. This deviation from the second law is found on finite time scales for small (or mesoscopic) systems and vanishes in the thermodynamic limit. Gallavotti and Cohen provided a proof of the FT for time-reversible Anosov systems \cite{Gallavotti_JSP:1995,Gallavotti_PRL:1995}. Dewar derived the FT based on a maximum entropy production principle \cite{Dewar:2003}.

The study is guided by the steady state Fluctuation Theorem (see e.g. \cite{Evans:2002})
\begin{equation} \label{FTsigma}
	\lim_{\tau \rightarrow \infty} 
	\frac{1}{\tau}
	\ln    
	\frac{P(p_\tau = A)}
     		{P(p_\tau = -A)}
     	=  \sigma_{+} A 
\end{equation}
for the ratio $p_\tau =\sigma_\tau/\sigma_{+}$ of  the time averages $ \sigma_\tau $ of the entropy production $\sigma$ in $\tau$-windows (beyond the relaxation to the steady state) and the long term mean $\sigma_{+}$.
%
%
The FT can be derived for the so-called dissipation function defined in phase space which needs identification with a macroscopic observable  \cite{Jepps_Rondoni:2016}.  In the following we will use the often used notion fluctuation relation (FR) for (\ref{FTsigma}).


The FT has been observed in a large number of laboratory and numerical experiments using different observables. In experiments the relation (\ref{FTsigma}) is valid for time scales $\tau$ well above characteristic time scales. Rayleigh-B\'{e}nard convection was studied by \cite{Shang:2005} for the local entropy production as observable. In numerical experiments of thermal convection \cite{Zonta:2016} analyzed the work term along Lagrangian paths as a representation of the entropy production rate. The work by the turbulent pressure force in two experiments was subjected to an FT analysis by \cite{Ciliberto_Garnier:2004}. The relation (\ref{FTsigma}) was found with modified slopes depending on the chosen time window and the impact of a new reference state was briefly considered. The injected power was used as an observable in different physical systems including the GOY turbulence shell model \cite{Aumaitre:2001}. In experiments with the model PUMA finite time Lyapunov exponents for the global circulation were observed with a frequency compatible with the FT \cite{Schalge:2013}. In all these hydrodynamic experiments the  time reversibility as a condition for the validity of the FT is not satisfied.


Our aim is two-fold: First we determine the distributions of the energy input and the currents. We consider the generalized Gumbel distribution and the Generalized extreme value (GEV) distribution. In the second step we attempt to assess the fluctuation relation. Thus, our approach is closely related to  \cite{Falcon:2008} on wave turbulence and to \cite{Falcon:2009}  on an electric circuit. In both studies the FT could not be verified. A major problem in our LEC data is  the lack of a reference state and the sparsity of negative data in the global averages. Therefore, we test shifts of the currents to two reference states, the GEV location parameter and the mean. 


The paper is organized as follows: The model is described in Section \ref{Sec_Model}  and the Lorenz energy cycle is defined in Section \ref{Sec_LEC}. The results for the densities are in Section \ref{Sec_Current_densities} and for the fluctuation ratios in Section \ref{Sec_FR}. A Summary and Discussion is included in Section \ref{Sec_Sum_Disc}.

\section{Global Circulation Model}  \label{Sec_Model}

To determine energy currents we use the model PUMA (Portable University Model of the Atmosphere, University of Hamburg) \cite{Fraedrich:2005,Fraedrich:2012}), a hydrostatic global atmospheric model based on the primitive equations on the sphere. The dynamical variables are vorticity, horizontal divergence, temperature and the logarithm of the surface pressure. The set of equations is
\begin{eqnarray}
\partial_t \xi
& = &
s^2
\partial_\lambda  {\cal F}_v
-\partial_\mu {\cal F}_u
-\frac{1}{\tau_f} \zeta
-K\nabla^{8} \zeta,
\\  \label{puma:a}
%
\partial_t D
& =& s^2
\partial_\lambda
{\cal F}_u
+\partial_\mu {\cal F}_v
-\nabla^2
  [         \frac{s^2}{2} (U^2 + V^2)
\\  \nonumber
&+& \Phi
 +  
\bar{T} \ln p_s  
]
 - 
\frac{1}{\tau_f} D
-K \nabla^{8} D, 
\\  \label{puma:b}
\partial_t T'
&=&-s^2
\partial_\lambda
(U T')
-\partial_\mu
(V T')
+D T'
-\dot{\sigma}\frac{\partial T}
{\partial \sigma}
\\  \nonumber
&+&\kappa \frac{T\omega}{p}
+\frac{1}{\tau_c}(T_R-T)
-K\nabla^{8} T',
\\    \label{puma:c}
\partial_t \ln p_s
&=&
-s^2 U
\partial_\lambda
\ln p_s
-V \partial_\mu 
\ln p_s -D
-\frac{\partial \dot{\sigma}}
{\partial\sigma},
\\
\label{puma:d}
%
\frac{\partial \Phi}{\partial \ln \sigma}
&=& -T,
\label{puma:e}
\end{eqnarray}
with $\mu=\sin \phi$ and $s^2 =1/(1- \mu^2)$. The variables $\zeta$ and $\xi$ denote absolute and relative vorticity, $D$ is the horizontal divergence and $p_s$ the surface pressure. The temperature $T$ is divided into the background, $\bar{T}$, and the anomaly, $T'$. Spherical coordinates are $\lambda$ and $\phi$ for longitude and latitude. $\Phi$ is the geopotential, $\kappa$ the adiabatic coefficient, and $\omega$  the vertical velocity. We use the abbreviations $U = u~ \cos \phi$ and $V = v~ \cos \phi$ for the zonal and meridional velocities $u$, $v$,  and the fluxes ${\cal F}_u=V\zeta- \dot{\sigma} \partial U/ \partial\sigma -T' \partial \ln p_s/ \partial\lambda $ and $ {\cal F}_v=-U\zeta-\dot{\sigma} \partial V/\partial\sigma -T's^{-2} \partial \ln p_s/\partial \sin \phi$. The vertical coordinate is divided into equally spaced $\sigma$-levels, $\sigma=p/p_s$, with the pressure $p$ and the surface pressure $p_s$.

A stationary state is maintained by driving the model towards a constant temperature profile (Newtonian cooling) with a prescribed equator-to-pole gradient. This means that a term $(T_{R}-T)/\tau_c$ is added to the temperature equation, where $\tau_c$ is the heating/cooling time scale, $T$ denotes the actual model temperature and $T_R$ refers to the prescribed reference temperature.  Dissipation is formulated as Rayleigh friction active in the boundary layer, i.e., terms $-\zeta/\tau_f$ and $-D/\tau_f$ are added to the equations for vorticity and divergence, where $\tau_f \approx 30$ days is the friction time scale.
Hyperdiffusion ($\propto K\nabla^{8}$) with a coefficient $K$ accounts for subscale processes and numerical stability.

The horizontal resolution is given by the total spherical wave number $\ell = 21$ with a triangle truncation and the vertical resolution is 10 vertical levels. The equations are numerically solved using the spectral transform method \cite{Orszag70}: Linear terms are evaluated in the spectral domain while nonlinear products are calculated in grid point space. In this configuration the model has  O($10^5$)  degrees of freedom. The model is integrated by a leap-frog method with a time step of 15 min. Orography is not specified and no external
variability like annual or daily cycle is imposed. The model is driven towards a mean state close to the observations.

\section{Lorenz Energy Cycle} \label{Sec_LEC}

\begin{figure}[h]
\includegraphics[angle=90, width=0.6\textwidth]{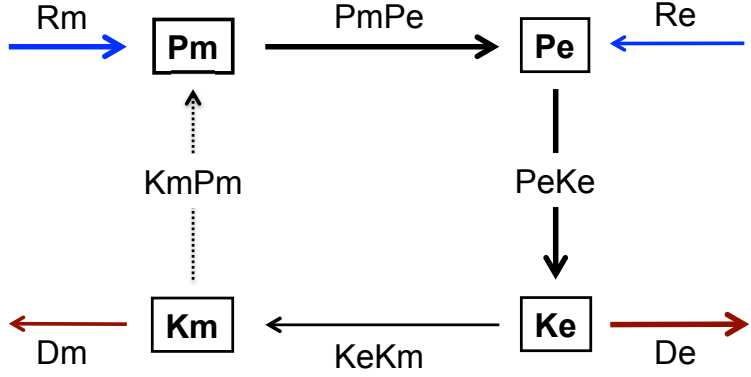}  
\caption{\label{Fig1} 
Lorenz energy cycle with energy compartments (boxes),  and energy currents (arrows). Available potential energy is P and kinetic energy K, zonal means are 'm'  and eddies 'e'. Forcing (R) is blue, dissipation (D) red, internal conversions black; intense currents are denoted by thick arrows, moderate thin, weak dotted.
}
\end{figure}

The atmospheric Lorenz  energy cycle (LEC) \cite{Lorenz:1955}  describes the general circulation from a perspective that emphasizes energy transformations, i.e., how the incoming solar radiation generates potential energy that is transferred to kinetic energy and finally lost to frictional dissipation (Fig. \ref{Fig1}). The LEC distinguishes the  zonal mean and deviations thereof. These so-called eddies can be identified with synoptic cyclones and anticyclones, with a length-scale of  thousand kilometers and a time-scale of several days; they play an important role in the atmospheric energy cycle. An early assessment of the LEC can be found in \cite{Peixoto:1974}, for a recent analysis in re-analysis data NCEP2 and ERA40 see  \cite{Li:2007}. The characteristics of the global atmospheric energy cycle are useful for the validation of general circulation; it is expected that the Lorenz energy cycle changes in a  warmer climate \cite{Deckers_Storch:2010}.

We calculate the following terms in the energy cycle, expressions can be found in \cite{Lorenz:1955} or in the textbook \cite{Holton:2012}: The forcings of the zonal mean $\mbox{Rm}$ and  the eddy available potential energy  $\mbox{Re}$, the dissipation rates of zonal mean $\mbox{Dm}$ and the eddy kinetic energies $\mbox{De}$. Conversion rates are determined between the zonal means of the kinetic and the available potential energies $\mbox{KmPm}$, the zonal mean and eddy available potential energies $\mbox{PmPe}$, eddy available potential and  kinetic energies $\mbox{PeKe}$, and eddy and zonal mean kinetic energies $\mbox{KeKm}$. 

The model was run for 1000 years and the LEC currents are determined as global means on a daily basis. For the interpretation it is relevant that the model is dry without convection and latent heat release. This means that there is no direct forcing on synoptic cycles and the mean of $\mbox{Re}$ is negative due to the damping effect of the zonal mean forcing. The reason is that the adjustment to the zonally constant temperature  attenuates the eddies and the model has no hydrological cycle and thus no latent heat release which could force the eddies.  

Note that in nature this damping term is caused by radiation and also present in complex models.  If a hydrological cycle with latent heat release is included this damping is compensated and the forcing $\mbox{Re}$ in observational data  has a positive mean \cite{Li:2007}. Please note that forcing and dissipative terms are only available indirectly in data. 

The present analysis faces two major problems: There is no reference state and large scale diffusivities or conductivities are unknown. This contrasts with Rayleigh-B\'{e}nard convection where a conductive state can be defined. Furthermore, there are few negative values due to the global averaging in the LEC currents. Therefore, we test the impact of shifting the currents to reference states.

\section{Current densities} \label{Sec_Current_densities}

\begin{figure}[h]
\includegraphics[angle=-90, width=0.6\textwidth]{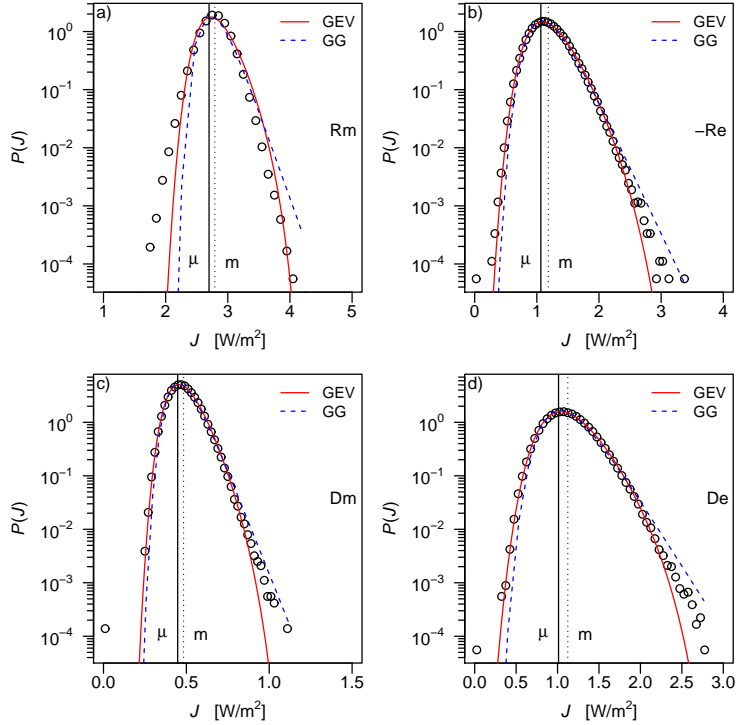} 
\caption{\label{Fig2} 
Normalized histograms of energy input and dissipation:
a) zonal mean forcing (injected power) $\mbox{Rm}$,
b) negative eddy forcing $-\mbox{Re}$, 
c) zonal mean dissipation $\mbox{Dm}$,
d) eddy dissipation $\mbox{De}$.
The solid (red) line is a GEV fit and the dashed (blue) line a generalized Gumbel (GG) fit.
The vertical lines indicate the GEV-location parameters $\mu$ (solid, black) and the means $m$ (dotted, black).
}
\end{figure}

\begin{figure}[h]
\includegraphics[angle=-90, width=0.6\textwidth]{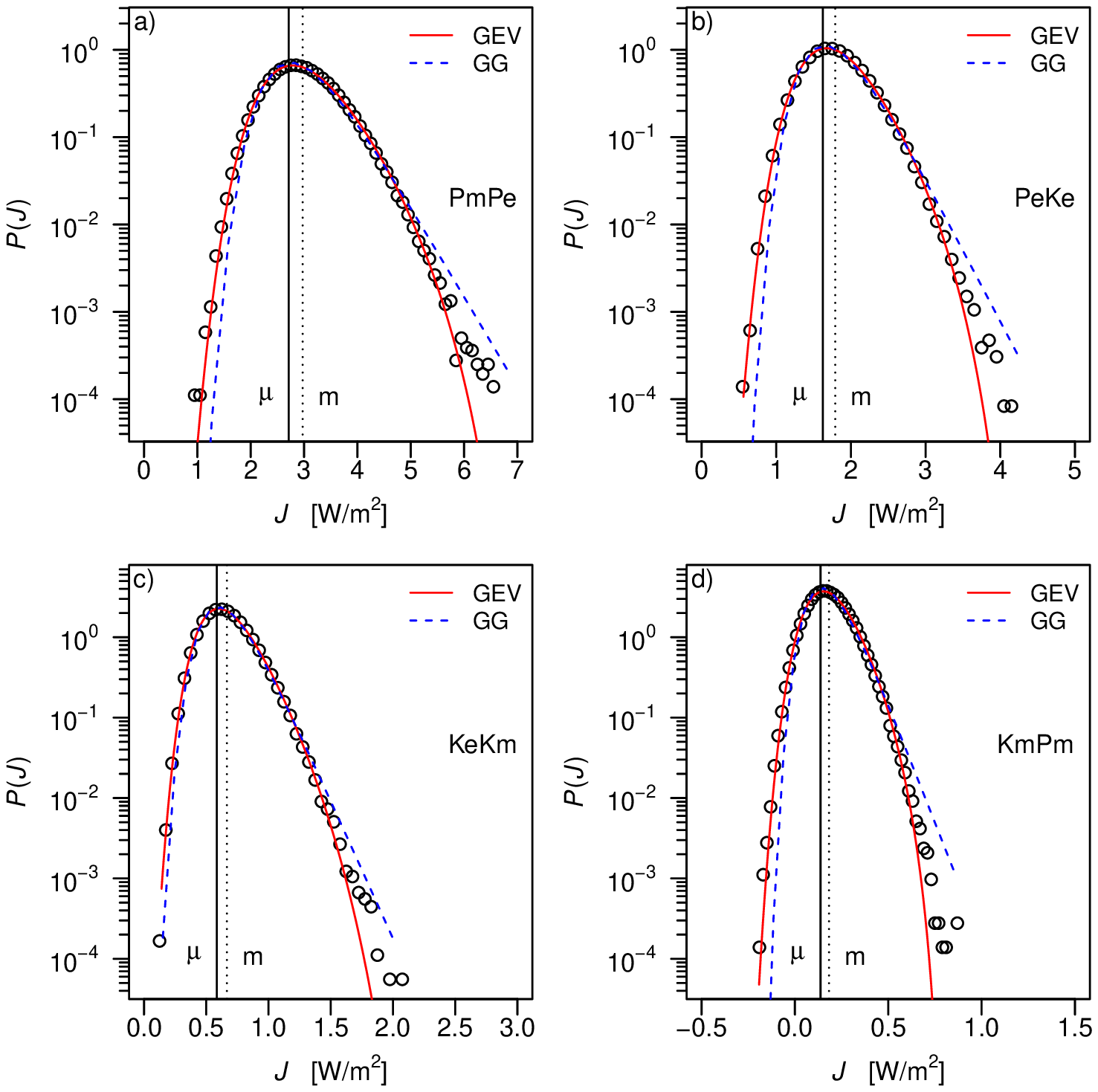} 
\caption{\label{Fig3} 
Normalized histograms of internal currents (conversions):
a) zonal mean to eddy available potential energy $\mbox{PmPe}$,
b) eddy available potential energy to eddy kinetic energy $\mbox{PeKe}$,
c) eddy kinetic energy to zonal mean kinetic energy $\mbox{KeKm}$,
d) zonal mean kinetic energy to zonal mean potential energy $\mbox{KmPm}$. The fits are as in Fig. \ref{Fig2}.
}
\end{figure}


The frequency distributions of the forcing terms  and the dissipative terms in the LEC are shown in Fig. \ref{Fig2}. For the eddy forcing the negative values are included, $-\mbox{Re}$, since the zonal mean forcing damps the eddies. The dissipative terms are split in the zonal mean part $\mbox{Dm}$ and the eddy part $\mbox{De}$.
The means of the internal currents are 
$\mbox{PmPe}$: 2.97,
$\mbox{PeKe}$: 1.79,
$\mbox{KeKm}$: 0.66,
$\mbox{KmPm}$: 0.18, 
and the means of the external currents are
$\mbox{Rm}$: 2.79,
$\mbox{Re}$: -1.18,
$\mbox{Dm}$: 0.48,
$\mbox{De}$: 1.12 (all values $\mbox{W/m}^2)$.
Note that the sign of the weak current $\mbox{KmPm}$ is unclear in observations \cite{Li:2007}.

The distributions can be approximated by the generalized Gumbel (GG) and the Generalized Extreme Value (GEV) distribution. In fits to the fluctuations of global quantities in correlated systems the generalized Gumbel distribution has been used (see \cite{Bertin:2005} and references therein). The density of the generalized Gumbel distribution is
\begin{equation}\label{GG-pdf}
	G_a(x) 
	= 
	\frac{\theta_a a^a}{\Gamma(a)} 
	\exp\{
	-[\theta_a(x+\nu_a) + e^{-\theta_a(x+\nu_a)}
	]
	\}
\end{equation}
with 
\begin{equation}\label{GG-pdf-2}
	\theta_a^2 
	= \frac{d^2 \ln \Gamma}{da^2}, \quad 
	\nu_a = \frac{1}{\theta_a}\left(\ln a - \frac{d \ln \Gamma}{da}
	\right)
\end{equation}

The GEV probability density is
\begin{equation}\label{GEV-pdf}
	f(z) = (1/s) (1+ \xi z)^{-1-1/\xi}, \quad z=(x-\mu)/s,
\end{equation}
with the location parameter $\mu$, the scale $s$, and the shape parameter $\xi$. For a vanishing shape parameter $\xi$ the GEV distribution reduces to the Gumbel distribution. The shape parameters $\xi$ of the currents in the Figs. \ref{Fig2} and \ref{Fig3} are in the range $\xi \approx -0.2, \dots, -0.1$. The skewness of the currents is positive and roughly $0.5$. 

As injected power in our model we consider the zonal mean forcing $\mbox{Rm}$ of the available potential energy. The forcing of the eddies $\mbox{Re}$ is not considered since it damps eddies and has a negative mean. Friction takes place mostly in the lowest levels which represent the atmospheric boundary layer, while the upper troposphere is only subject to hyper-diffusion (this is the reason for the meteorological notion of the so-called free atmosphere). 

The forcing of the zonal mean potential energy, which is the energy input in the present simulation, is used to quantify the entropy production in the non-equilibrium system,
\begin{equation}
     \sigma =  \mbox{Rm},
\end{equation}
with the  long term mean defined by $\sigma_{+} = \langle \sigma \rangle$. The reason for this choice is that the eddy forcing $\mbox{Re}$ acts as a  dissipation since the relaxation to a zonal mean temperature attenuates eddies. Note that the means satisfy  
\begin{equation} \label{means}
	\langle \mbox{Rm} \rangle 
	\approx 
	\langle \mbox{Dm} + \mbox{De} - \mbox{Re} \rangle.
\end{equation}
Thus $\mbox{Re}$ should be added to $\mbox{Dm} + \mbox{De}$ and the common definition of an entropy production in terms of friction is not possible here. 

\section{Fluctuation ratio} \label{Sec_FR}

The ratio of negative to positive values in the currents is low and insufficient for an analysis of the fluctuation ratio. Therefore, we test shifts of the currents to reference states.  In the following we consider three reference states for the currents $J_R$:  (i) the location parameter $\mu$ of the fitted GEV distribution, (ii) the mean of each current, and (iii) the mode (pdf-maximum) of each current. 

The currents are transformed to anomalies
\begin{equation} \label{Jano}
     J^\prime = J - J_R 
\end{equation}
with the reference state $J_R$.

The anomalies are averaged in windows with length $\tau$
\begin{equation} \label{Jano1tau}
     J^\prime_\tau 
     = \frac{1}{\tau} \int_t^{t+\tau} 
     	J^\prime(t^\prime) \mbox{d}t^\prime 
\end{equation}
All averaged current anomalies $J^\prime_\tau$ are nondimensionalized by the long term mean entropy production $ \sigma_{+} = \langle Rm \rangle$

\begin{equation} \label{pJsigma}
     p_\tau=J^\prime_\tau/\sigma_{+}.
\end{equation}
The fluctuation ratio is determined for the anomaly ratios $p$  for the entropy production $\sigma$, and all other currents
\begin{equation} \label{FRalpha}
	\frac{1}{\tau}  
	\ln
	\frac{P(p_\tau = A)}
     		{P(p_\tau = -A)}
     	=  \beta A \sigma_{+}, 
\end{equation}
where we have introduced a slope $\beta$. The normalized time scale $\tilde{\tau}$ is obtained by a typical correlation time of all currents, $\tau_c = 5$ days, 
\begin{equation} \label{tautilde}
     \tilde{\tau} = \tau/\tau_c.
\end{equation}
So far it is unclear how to interpret the slopes $\beta$ as inverse turbulent temperatures (see e.g. \cite{Zonta:2016}). 

\subsubsection{Location parameter as reference state}

\begin{figure}[h]
\includegraphics[angle=-90, width=0.6\textwidth]{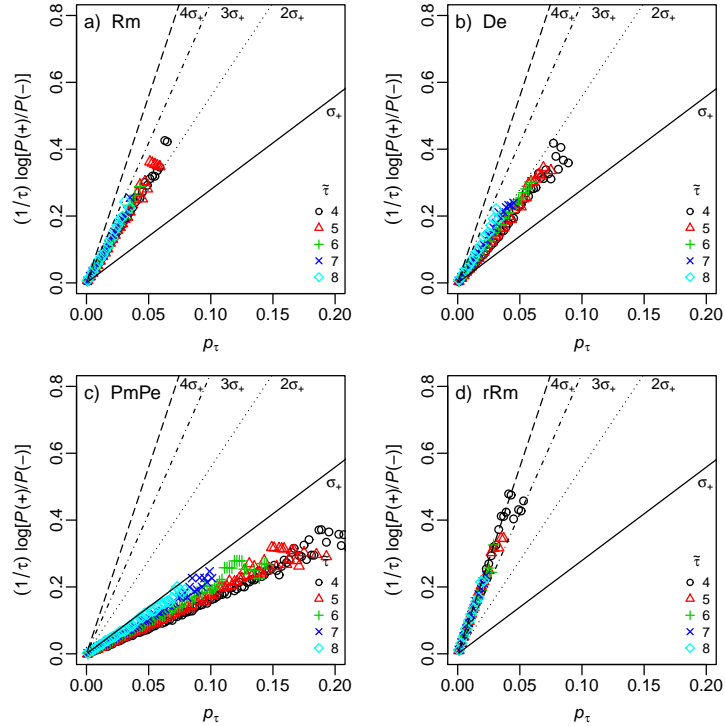} 
\caption{\label{Fig4} 
Fluctuation ratio for the shift to the location parameters:
a) injected power $\sigma =\mbox{Rm}$ (defined as the entropy production),
b) eddy dissipation $\mbox{De}$, 
c) conversion of zonally averaged potential to eddy potential energy $\mbox{PmPe}$,
d)  i.i.d. GEV random variates $\mbox{rRm}$ with the  distribution of $\mbox{Rm}$. 
}
\end{figure}

The first reference state is the location parameter defined for each current by $J_R=\mu_J$, determined by a GEV fit to  $J$. In Fig. \ref{Fig4} the results for (a) the injected power $\mbox{Rm}$,  (b) the eddy dissipation $\mbox{De}$, (c) the current $\mbox{PmPe}$ and (d)  surrogate data $\mbox{rRm}$ are shown. The current $\mbox{PmPe}$ is used as an example to represent the currents in the LEC. The surrogate data $\mbox{rRm}$ are independent random variates with a GEV distribution and parameters determined by a fit to  $\mbox{Rm}$ (injected power and entropy production $\sigma$). This data is added to extract the impact of the distribution independent of the correlations. Unfortunately, a robust quantitative estimation of the slopes is not possible, thus we refer to the slopes indicated in Fig. \ref{Fig4}. 

The fluctuation ratios in ({\ref{FRalpha}) for the injected power $\mbox{Rm}$ are linear with slopes between  $2 \sigma_{+}$ and $3 \sigma_{+}$. The eddy dissipation $\mbox{De}$ reveals linear slopes of the order of $2 \sigma_{+}$. The internal conversion  $\mbox{PmPe}$ is linear with slopes below $\sigma_{+}$. The surrogate data $\mbox{rRm}$ shows slopes  $\approx 4 \sigma_{+}$ independent of the average time $\tilde{\tau}$ since the data are uncorrelated. The slope in the injected power does not reach this value even for the longest times analyzed.

\begin{figure}[h]
\includegraphics[angle=-90, width=0.6\textwidth]{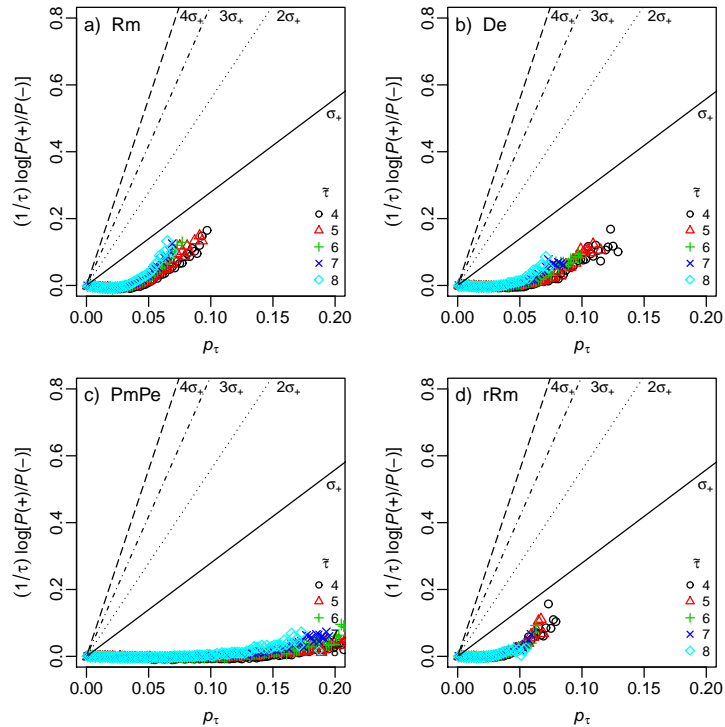} 
\caption{\label{Fig5} 
Fluctuation ratio for the shift to the means of the current distributions:
a) injected power $\sigma =\mbox{Rm}$,
b) eddy dissipation $\mbox{De}$, 
c) current zonally averaged potential to zonally averaged kinetic energy $\mbox{PmPe}$,
d)  i.i.d. GEV random variates $\mbox{rRm}$ with the  distribution of $\mbox{Rm}$. 
}
\end{figure}

\subsubsection{Mean as reference state}

For an assessment of the location parameter as the reference state we compare it to the mean of each current which could be considered as a first and nearby choice to increase the number of negative values. In Fig. \ref{Fig5} the results for the same currents as in Fig. \ref{Fig4} are shown. Obviously the fluctuation ratios are far from being linear. However, for large averaging times the slopes bend towards the slopes obtained for the location parameter (Fig. \ref{Fig4}). 

\subsubsection{Mode as reference state}

As a further alternative for a  reference state we have tested the mode $M_J$ (the maximum of the pdf) for each current, $J_R=M_J$. The choice of the mode can be motivated by the observation of cusps in the distributions of fluxes in laboratory experiments, e.g. for the local convective heat flux in  Rayleigh-B\'{e}nard convection \cite{Shang:2005} and the injected power in wave turbulence \cite{Falcon:2008}. The results for the mode (not shown) are close to the results for the location parameter in Fig. \ref{Fig4}. The reason is that the mode $M$ of the GEV distribution is 
\begin{equation}\label{Mode-GEV}
 	M=\mu + s ((1+ \xi)^{-\xi}-1)/\xi, 
\end{equation}
which is close to the location parameter $\mu$, since $M \approx \mu - s \xi$, for small shape parameters as found here ($-0.2, \dots, -0.1$). A clear advantage of the mode is that it can be estimated without an assumption on the distribution.

\section{Summary and Discussion} \label{Sec_Sum_Disc}

We have analyzed the atmospheric energy cycle defined by Lorenz \cite{Lorenz:1955} for the zonal mean and the eddy parts of the available potential and the kinetic energies. The LEC constitutes a network of energy currents in the atmosphere including the injected power and  the dissipated energies. Thus the LEC can be considered as an atmospheric energy cascade model. The means in the LEC constitute the climate from a dynamical point of view and the fluctuations are related to climate variability. Note that the properties of the climatological LEC are known from the output of models only, mainly in the reanalysis datasets ERA and NCEP \cite{Li:2007}.  

The LEC data used here is produced with the atmospheric model PUMA  (Portable University Model of the Atmosphere, University of Hamburg) which is a dynamical core based on the hydrostatic primitive equations \cite{Fraedrich:2005}. The model uses linear forcing and friction for unresolved processes.  The forcing is chosen to obtain a steady state close to the observations.  The model was run for 1000 years and the LEC data consists of daily global averages. 

The LEC current distributions can be approximated by the generalized Gumbel distribution and the Generalized Extreme Value (GEV) distribution. As \cite{Bertin:2005} pointed out that the frequently found generalized Gumbel distribution can be derived for correlated systems.

The Fluctuation Theorem (or fluctuation relation) relates the probabilities of negative and positive entropy productions in non-equilibrium physical systems. Here, the aim is to use the steady state version to constrain the  distribution of current anomalies in the LEC. The entropy production $\sigma$ is defined as the injected power. For the analysis of the FR in the atmosphere it is unfavorable that there is no reference state and that the globally averaged LEC reveals few negative values.  To overcome both problems we shift the currents to reference states. For the reference states we use the location parameter obtained by a fit of the Generalized Extreme Value (GEV) distribution, and the mean. A nondimensional time scale is defined by $\tilde{\tau} = \tau/\tau_c$, where $\tau_c=5$ days is a typical correlation time scale of the currents. 

We define anomalies for the currents with respect to the two reference states and nondimensionalize them with the long term mean $\sigma_{+}$ of the entropy production. For the location parameter reference state, the currents follow fluctuation relations with linear slopes in the range $\sigma_{+}  \dots 4 \sigma_{+}$.  In the analysis, a surrogate data test is included which uses i.i.d. random variates distributed as the entropy production. For the location parameter the FR for this data has a slope $\approx 4 \sigma_{+}$.  

We conclude with remarks on the applicability of the FT. 
(i) On the local FT: In numerical models any local variable is averaged due to the finite model resolutions. Thus local variables are not accessible. 
%
%
(ii) On time reversibility: A common notion in meteorology is the so-called {\it free atmosphere} above the boundary layer (the lowest hundreds of meters were friction takes place) \cite{Holton:2012}. Therefore, it might be reasonable to assume that the atmosphere is approximately time-reversible on  the corresponding time scales. 


Since no physical constraints for the distributions of atmospheric energy currents are known we expect that our findings might be useful for model assessment, global warming studies \cite{Deckers_Storch:2010}, and possibly the behavior of extremes (see for example \cite{Merhav:2010,Gundermann:2014}).

\section*{Acknowledgement}

We like to thank for the support by the DFG Transregio project TRR181 ("Energy Transfers in Atmosphere and Ocean").

\clearpage

\bibliographystyle{unsrt}

\bibliography{Blender-Lorenz}

\end{document}